# AI-GENERATED INTERACTIVE FICTION FOR EDUCATIONAL USE: A PILOT STUDY OF PERCEIVED COMPREHENSIBILITY, COHERENCE, AND ENGAGEMENT

**F. Rogosch, A. Schrader**

*University of Lübeck (GERMANY)*

## Abstract

Generative artificial intelligence (AI) can produce educational content at scale, including interactive and narrative learning experiences, but technical generation alone is not sufficient: scenarios that are confusing, narratively inconsistent, or unengaging are unlikely to be useful in practice. This paper presents a pilot user-centred evaluation of AI-generated interactive fiction (IF) for educational use in higher education. Using a previously described domain-agnostic pipeline and a shared STEM content base, we generated a controlled pool of scenarios and asked participants ($N$ = 22, STEM higher-education) to play one generated episode and rate it on narrative clarity, story-content coherence, engagement, and length acceptance. A free-text prompt captured open feedback. Narrative clarity and length acceptance were rated positively, engagement sat near the neutral mid-point of the scale, and story-content coherence was the weakest dimension by a clear margin. Qualitative feedback points to quiz integration as the bottleneck. Artificial in-fiction motivation for quiz prompts and abrupt setting changes were reported. Feedback also pointed to missing story-level consequences for wrong answers. From these observations, we derive concrete design implications that can inform larger follow-up studies, including later work on learning effectiveness.

## 1 INTRODUCTION

Generative artificial intelligence (AI) is making it easier to produce educational content at scale, including interactive and narrative learning experiences [1], [2]. Interactive fiction (IF), short choice-driven text-based scenarios, is a promising format for game-based learning because it embeds educational content into branching narratives that learners explore through decision-making [3], [4]. AI-assisted generation could lower the well-documented authoring bottleneck of serious games in higher education [5], [1].

However, technical generation alone is not sufficient. Reviews of LLM-based game and narrative generation report that runnable outputs are within reach while narrative quality and the meaningful integration of game rules and story remain difficult [2], [6], with narrative coherence flagged as a recurring weakness [7], [8]. For educational use, the question is therefore not only whether such scenarios can be generated, but how they are experienced by the learners who interact with them.

This paper presents a pilot evaluation of AI-generated IF for educational use in higher education. It builds on prior work that introduced a domain-agnostic pipeline for automatically generating IF scenarios from structured educational content and evaluated the pipeline primarily from a generation-pipeline perspective (compilation, playability, and learning-goal fidelity) [9]. The present study shifts the focus from the pipeline to the perceived quality of the generated scenarios. Using the same pipeline in one STEM sub-domain, we generated a controlled stimulus pool and asked STEM higher-education participants to play one scenario and rate it on three perceived-quality dimensions: narrative clarity, story–content coherence, and engagement. A free-text prompt captured strengths, points of confusion, and weaknesses that structured ratings may miss. Learning effectiveness is deliberately outside the scope of this pilot. Instead, we check whether the generated scenarios meet the basic perceived-quality requirements a learning-outcome study would depend on: that learners can follow the story, perceive the educational content as integrated into it rather than appended as a separate quiz [10], and are engaged enough to play through [11]. The broader research goal these stimuli serve is to make such scenarios deployable as short, spatially situated learning episodes in higher-education spaces, in the spirit of ambient serious games [12]. The perceived-quality requirements tested here are a prerequisite for that direction.

The contribution of this paper is twofold. First, we report an early user-centred evaluation of how IF learning scenarios produced by this specific pipeline configuration are perceived by higher-education participants, pairing automated generation metrics with structured ratings and qualitative feedback on a controlled stimulus pool. Second, we translate the results into concrete design implications for AI-generated IF in education, highlighting where the format is already usable and where additional stimulus engineering is needed before moving to larger confirmatory studies.

# 2 METHODOLOGY

This shift from generation success to perceived scenario quality requires explicit criteria for usable educational IF, beginning with intrinsic integration.

## 2.1 Background

Serious games are digital games developed for purposes beyond entertainment [11]. A key design principle for educational games is intrinsic integration: learning content should not be layered as a separate quiz on top of a narrative but woven into the decisions and causal structure of the story [10]. This motivates our focus on perceived story–content coherence as a central evaluation dimension alongside narrative comprehension and engagement.

## 2.2 Stimulus Generation

Study stimuli were produced by SINE, a previously described domain-agnostic pipeline that generates choice-based IF scenarios from structured educational content using an open-weight LLM, with deterministic validation and a repair stage [9]. The pipeline was kept unchanged; only the inference backend was moved to Qwen3 14B [13] accessed via the OpenRouter cloud API [14] to make generation tractable. Prompt template, generation strategy, and validation were fixed after the refinement phase below and held constant across all runs.

Controlled content base. We defined a fixed set of 20 educational seeds from a single STEM sub-domain (media technology) and used them as a controlled content base, rather than drawing from a heterogeneous existing output set. Each seed specifies the learning content at a comparable level of granularity, so that differences between scenarios primarily reflect narrative generation rather than input heterogeneity. The full seed set is archived alongside the generated stimulus pool and the frozen prompt configuration in a public repository alongside the anonymised response data.

Structured prompt refinement. Early pilot generations passed the pipeline's technical gates but showed defects visible to a human player (abrupt endings, dominant shortcut paths, quiz prompts being placed on in-story terminals or displays without being integrated into the scene logic). We therefore ran a structured refinement phase on a held-out seed set, iterating the prompt configuration against minimum-quality gates covering playability, narrative adequacy, educational integration, target play duration of 5–10 minutes, and absence of shortcut-only completions. Hard criteria were enforced by the validator [9]; soft criteria were judged by the first author on a small story sample. Refinement targeted general structural quality, not the items used later in the questionnaire. After the final iteration, the configuration was frozen and all study stimuli were generated in a single batch.

Final stimulus pool. Generation used 20 seeds with three repetitions per seed. The seeds covered five topics within the media technology sub-domain: sampling, quantization, lossless compression, lossy compression, and JPEG image compression. After automated playability and validation filtering, 48 scenarios remained (two to three per topic) and formed the study pool, with each participant randomly assigned one scenario. Retaining the full pool preserves within-topic variation across narrative realizations and reduces overlap between participants.

## 2.3 Participants and Procedure

Participants were recruited via internal mailing lists at the authors' institution and personal contacts, targeting adults (≥ 18 years) with a STEM higher-education background, namely current students, staff members, or recent graduates of a STEM programme. The study was fully online and self-paced. After giving informed consent, each participant was randomly assigned one scenario from the pool and played it to completion in the browser (typical playtime 5–10 minutes), then completed the questionnaire (2–3 minutes). Stimuli were presented in English. Consent, instructions, and questionnaire were in German to reduce language-related measurement error in the predominantly

German-speaking cohort. Data were collected between 2026-04-09 and 2026-04-20. Figure 1 shows the scenario player during play.

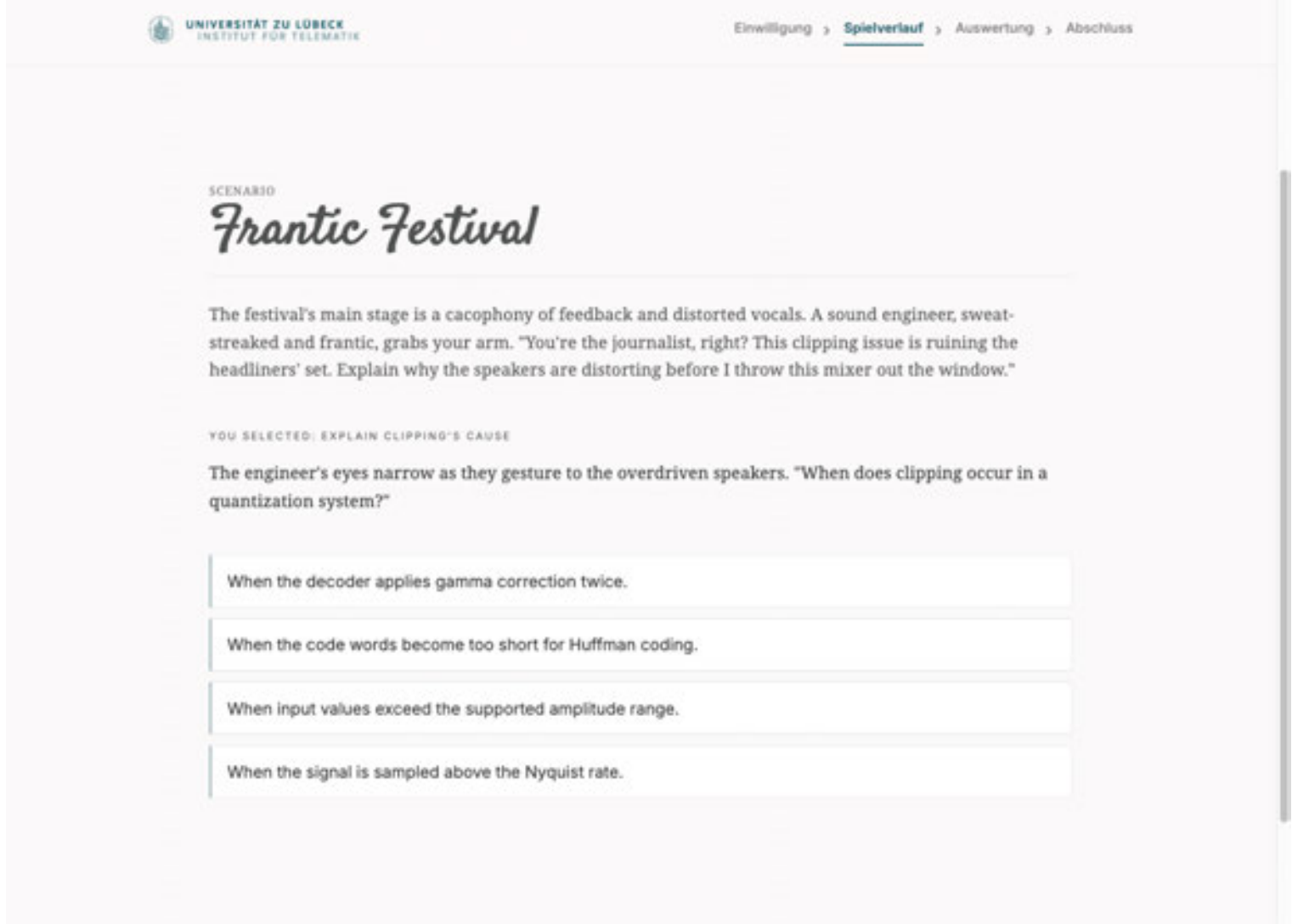

*Figure 1. Browser-based scenario player on a quiz beat: scenario title and story passage with the four available multiple-choice answers, study-flow breadcrumb in the top navigation.*

Inclusion required a STEM higher-education affiliation and a fully completed session (consent, scenario, and questionnaire). Incomplete sessions were not transmitted to the server. Post-hoc exclusions covered participants who selected "none of the above" for affiliation, internal test sessions, sessions before the recruitment start time, and repeat participations (to preserve between-subjects independence against novelty loss, demand characteristics, and carry-over effects). After these criteria were applied, $N = 22$ sessions entered the analysis (8 currently enrolled students, 11 university staff members, and 3 recent graduates).

## 2.4 Measures

We developed a compact questionnaire for the three target constructs, narrative clarity, story-content coherence, and engagement, plus a single length-acceptance item, following a theory-driven adaptation approach in line with established brief-scale guidelines [15], [16]. Rather than carrying over full original scales, we adapted individual items from the Narrative Engagement Scale [17], the Transportation Scale–Short Form [18], MEEGA+ [19], and IMI [20] to the IF context, complemented by theory-guided items for story-content coherence grounded in the intrinsic-integration principle [10]. All items were phrased as positively-worded statements; we omitted reverse-scored items, a common short-pilot trade-off [15], [16]. The questionnaire was kept short to fit the 5–10 minute time budget of the online study, accepting a less complete psychometric coverage than longer established scales would offer. The final instrument comprises ten positively-worded Likert items on a five-point scale from “strongly disagree” (1) to “strongly agree” (5), grouped into four subscales: Narrative Clarity (2 items), Story–Content Coherence (4 items), Engagement (3 items), and Length Acceptance (1 item), followed by four control items and one open-ended prompt. The nine Likert items for the three subscales and their adaptation from the host instruments are documented in the public item pool [21]; the length-acceptance item, control items, and open-ended prompt were added for the present study and are reported here. The length-acceptance item read: “The length of the scenario was appropriate.” (presented in German: “Die Länge des Szenarios war für mich angemessen.”).

Control items, presented in German, covered affiliation („Welchen Bezug hast du zu einer Hochschule?“ — Eingeschrieben / Mitarbeitende:r / Alumni / Keiner), number of prior runs including the current one („Wie oft hast du diese Studie bisher gestartet (einschließlich dieses Durchlaufs)?“ — 1-mal / 2-mal / 3-mal oder häufiger / Weiß nicht), subject-matter familiarity („Wie gut kennst du dich mit Medientechnik (Grundbegriffe/Konzepte) aus?“, 5-point „gar nicht“ bis „sehr gut“), and perceived English-language barrier of the scenario („Die englische Sprache des Spiels hat es mir erschwert, das Szenario zu verstehen.“, 5-point). The open-ended prompt, presented in German, asked: „Was war besonders gut, verwirrend oder verbesserungswürdig (Story, Entscheidungen, Einbettung der Fachinhalte, Länge/Tempo)?“ Gameplay telemetry captured play duration, the first-try-correct rate across quiz interactions, and the total number of wrong clicks.

## 2.5 Analysis Plan

Analyses are primarily descriptive. For each item and each sub-dimension, we report means, standard deviations, medians, response-category distributions, and, at the sub-dimension level, 95% confidence intervals for the mean. Sub-dimension scores are computed as the arithmetic mean of their items. Internal consistency is reported per sub-dimension: Cronbach's α for the three-item engagement and four-item coherence subscales, Spearman-Brown for the two-item clarity subscale, and n/a for the single-item length-acceptance subscale [15], [16]. To check whether self-report was driven by quiz performance rather than perceived quality, we additionally computed Spearman rank correlations between each sub-dimension and the gameplay metrics. As a sensitivity check, we also correlated each sub-dimension with the self-reported English-language barrier. Open-ended responses were inductively grouped into recurring themes by a single rater across multiple passes, and linked back to the quantitative scores of the corresponding participants with short representative quotations. The themes illustrate the quantitative findings rather than making claims of their own; the single-rater setup is acknowledged as a limitation in Section 4.3. Given $N = 22$, any inferential indicators (correlations, group contrasts) are reported as exploratory only and are not interpreted as confirmatory tests. The anonymised dataset underlying all analyses reported here is publicly available [22].

# 3 RESULTS

The descriptive results indicate which perceived-quality requirements the generated scenarios meet, and where they remain insufficient.

## 3.1 Sample Composition

The 22 included participants played 19 distinct scenario files (three files were played by two participants each). Self-rated subject-matter familiarity was moderate (M = 2.68, SD = 0.99) and the perceived English-language barrier of the scenario was low (M = 1.73, SD = 0.94), both on the same five-point scale. All included sessions are first-time participations. Two repeat-participation sessions were removed by the exclusion criterion described in Section 2.3.

## 3.2 Perceived Quality

Table 2 summarizes the four sub-dimension scores, with 95% confidence intervals and internal-consistency estimates.

*Table 2. Sub-dimension scores across N = 22 participants. k is the number of items. Rel. is Cronbach's α for k ≥ 3, Spearman-Brown for k = 2, and n/a for the single-item length-acceptance subscale. Subscale scores are item means on the 5-point Likert scale, with t-based confidence intervals.*

| Subscale | k | Rel. | M | 95% CI | SD | Mdn | Min–Max |
|---|---|---|---|---|---|---|---|
| Narrative Clarity | 2 | 0.63 | 4.11 | [3.74, 4.49] | 0.84 | 4.25 | 2.50–5.00 |
| Story–Content Coherence | 4 | 0.87 | 2.92 | [2.47, 3.37] | 1.02 | 3.12 | 1.00–4.25 |
| Engagement | 3 | 0.82 | 3.08 | [2.58, 3.57] | 1.12 | 3.00 | 1.33–5.00 |
| Length Acceptance | 1 | — | 4.14 | [3.74, 4.53] | 0.89 | 4.00 | 2.00–5.00 |

Narrative clarity and length acceptance were rated positively (both means above 4.10), engagement hovered near the neutral mid-point with its 95% CI straddling it, and story–content coherence was clearly the weakest dimension: its scale CI's upper bound only just brushes the mid-point. Three of the four coherence items scored below the mid-point, as did the absorption item within engagement (M = 2.64); all other items scored above it. Internal consistency was acceptable-to-good for the multi-item subscales (Cronbach's α = 0.87 for coherence, α = 0.82 for engagement; Spearman-Brown = 0.63 for the two-item clarity scale). At the item level, responses on the coherence subscale are spread across the full scale with a clear mass below the neutral mid-point on all four items, echoed in the qualitative feedback (see Section 3.4).

## 3.3 Gameplay and Correlations with Self-Report

Play duration had a median of 5.7 minutes, in line with the intended 5–10 minute scope. One outlier session (54.6 minutes) most plausibly reflects a participant leaving the page open rather than continuous play and is treated descriptively. The first-try-correct rate across embedded quiz items was M = 0.71 (SD = 0.25). Spearman rank correlations between each sub-dimension and the gameplay metrics (duration, extra clicks, first-try-correct rate, total wrong clicks) were all small-to-moderate ($|\rho| \leq 0.38$) and none reached the uncorrected $p < 0.05$ threshold. The strongest tendency was a moderate negative association between engagement and extra clicks ($\rho = -0.38$, $p = 0.08$), intuitive but far from confirmatory at this sample size. As a sensitivity check on the English-language stimulus, correlations with the self-reported language barrier were negligible ($|\rho| \leq 0.10$ across all four sub-dimensions), so language issues do not plausibly explain the coherence result. Self-report on the four sub-dimensions is therefore largely decoupled from observable quiz performance, which is a useful precondition for interpreting the coherence result in content-related rather than frustration-related terms.

## 3.4 Qualitative Feedback

Ten of 22 participants (45%) left a free-text response. We inductively clustered the responses into ten recurring themes, which split into content-level criticisms (Themes 1–7), tooling and UX issues (Themes 8 and 10), and positive remarks (Theme 9). Table 3 lists the themes and their session frequencies.

*Table 3. Inductive themes from the ten free-text responses (N = 22).*
*Most responses contained more than one theme, so counts are session-level.*

| # | Theme | Sessions |
|---|---|---|
| 1 | Artificial in-fiction motivation for quiz prompts | 6 |
| 2 | Positive remarks on idea, enjoyment, or engagement | 5 |
| 3 | Abrupt location or character changes without narrative bridge | 2 |
| 4 | Wrong answer options too obvious or solvable by elimination | 2 |
| 5 | Pacing: context shifts too fast, or scenario too short | 2 |
| 6 | Redundant question repetition (pipeline generation issue) | 2 |
| 7 | No consequences for wrong answers (“click-through” possible) | 1 |
| 8 | Monotony: every choice is a learning item, never a story choice | 1 |
| 9 | Response-type monotony (only multiple-choice) | 1 |
| 10 | Language mismatch (German UI labels with English story) | 1 |

The dominant criticism is the artificial in-fiction motivation for the quiz prompts (Theme 2, six of ten responses): participants ask why the non-player characters would demand technical knowledge of the player at all, and several explicitly link this to a drop in perceived coherence. Two participants independently propose the same remedy: moving from realistic to surreal or fictional settings to free the scenario from real-world causal plausibility. A second, more severe cluster (Theme 3, two responses) involves abrupt location or character changes mid-scenario ("Suddenly we were teleported from a museum to a radio station, with new characters praising me who had not been there before."), co-occurring with the lowest coherence scores in the sample ($\leq 1.75$). A third concern (Theme 1, one response) is that wrong answers only produce a "try again" without story-level consequence, making "click-through" possible. This is directionally consistent with the engagement–extra-clicks correlation reported in Section 3.3, but as a single-participant observation should be treated as suggestive. Themes 8 and 10 (question repetition, German UI / English story mismatch) are surface-level pipeline and tooling issues outside the perceived-quality interpretation.

The quantitative and qualitative patterns align. Coherence-critical sessions do not come from the low-performance end of the sample: participants with poor quiz performance wrote constructive rather than dismissive feedback, and the two lowest coherence ratings in the sample were written by participants with mid-to-high first-try-correct rates (3/5 and 4/5) who identified specific setting-level inconsistencies regardless of their own performance. Positive remarks appear across the sample, including in sessions whose scores were otherwise mixed, suggesting that the IF format itself is accepted even where the current stimulus implementation is criticized.

# 4 DISCUSSION

Taken together, the ratings and comments point to a specific mismatch: the scenarios were comprehensible, but the educational questions were not consistently integrated into the story world.

## 4.1 Interpretation

The scenarios produced by the pipeline were perceived as comprehensible and of appropriate length, while engagement was marginal (the mid-point lies within the 95% confidence interval of the engagement mean). Story-content coherence, by contrast, is the clear bottleneck. Three of its four items fell below the neutral mid-point of the scale, the confidence interval of the scale mean lies almost entirely below it (with its upper bound just touching the mid-point), and the qualitative feedback converges on an identifiable source for this pattern: the narrative integration of the quiz prompts rather than the narrative surface itself. This interpretation is reinforced by the correlation analysis: coherence ratings were essentially independent of quiz performance, so low coherence scores are unlikely to reflect frustration over wrong answers. Nor are they plausibly driven by the English-language stimulus, given the small language-barrier correlation. Prior work has shown that LLM-generated game content can be coherent at the surface level but struggles when multiple generation constraints interact [6], [8]. Our finding extends this pattern to educational IF and localises it precisely at the seam between story and educational content. We read this bottleneck as an artefact of the specific pipeline configuration rather than as evidence about LLM-generated IF in general. The clarity, length, and positive free-text remarks confirm that the IF format itself is accepted at this scale.

## 4.2 Design Implications

The qualitative clusters and the interpretation above translate into two families of design targets, with clearly different evidentiary weight.

Participant-driven changes to the generated scenarios. First, the in-fiction motivation for why non-player characters ask the player for technical knowledge must be made explicit in the narrative (Theme 2, six of ten free-text responses). Participants independently suggested that surreal or fictional settings may be more productive than realistic ones because they remove the implicit contract that asks for real-world causal plausibility. Second, abrupt location or character changes are disproportionately damaging (Theme 3, two responses): they coincide with the lowest coherence scores in the sample and should be suppressed in the generation or repair stage. A further suggestive observation, based on a single participant (Theme 1) and consistent with the engagement–extra-clicks correlation, is that scenarios could attach a story-level consequence to wrong answers rather than only looping back with a "try again". Complementary but less urgent suggestions include mixing purely narrative choices with learning choices and varying the interaction format beyond multiple-choice. All of these are design-change hypotheses suggested by participant feedback, not tested interventions.

Pipeline-level changes suggested by stimulus preparation. Our finding has a direct consequence for the quiz-fidelity check: passing it does not predict perceived narrative integration. The implication is to replace verbatim fidelity with a semantic-equivalence check (e.g. an LLM judge with human review for ambiguous cases) and to strengthen the playability check from start-to-end reachability to traversal- or objective-coverage so shortcut completions no longer pass. Integrating these checks into the repair loop, rather than applying them only as a post-hoc filter, is the natural next step.

## 4.3 Limitations

The study is a small convenience-sample pilot in a single STEM sub-domain, with single-exposure sessions and deliberately no learning-outcome measure. At $N$ = 22, any correlations and group contrasts are exploratory, and the within-stimulus spread we observed, which was comparable in magnitude to the between-stimulus spread, prohibits reliable per-scenario rankings. We therefore report only at the sub-dimension level.

The stimuli were produced by a single pipeline configuration with one base model (Qwen3 14B) in one STEM sub-domain (media technology). Qwen3 14B was inherited from the preceding pipeline-evaluation study, in which it achieved the highest end-to-end success rate among the open-weight models tested [9]. An informal comparison generation with a substantially larger cloud-only model (Qwen3.6 Plus via OpenRouter) under the same prompt configuration produced visibly stronger narrative and question-integration behaviour, consistent with the capacity interpretation above. The findings are therefore specific to this pipeline-model combination, and cross-model and cross-domain generalisation is out of scope.

Recruitment via internal mailing lists and personal contacts may introduce a pro-technology self-selection and demand-characteristics bias, and a STEM-expert audience may be unusually sensitive to narrative-content mismatch in its own domain, so the coherence result may not transfer to non-expert learners. The questionnaire adapts individual items from established instruments and has not yet been subjected to a full psychometric validation, as is appropriate for this early pilot stage [16]. Internal-consistency estimates (Section 3.2) are promising for the multi-item subscales but the two-item clarity reliability of 0.63 is borderline. Qualitative coding was performed by a single rater with knowledge of the quantitative scores. This introduces a confirmation-bias risk that would be mitigated in a follow-up study by blinded double-coding with inter-rater agreement reporting.

### 4.4 Future Work

Three complementary strands of follow-up work suggest themselves. First, the pipeline-level changes outlined above need to be integrated into the repair loop and re-evaluated: stronger open-weight base models, softer semantic-equivalence quiz validation in the spirit of intrinsic integration [10], and traversal- or objective-coverage playability. Second, spatially situated, station-based delivery connects this format to ambient serious games [12]: once perceived-quality requirements are met on-screen, the same scenarios can be evaluated in a setting where each narrative beat is anchored to a physical location. Third, a dedicated learning-outcome study with a brief knowledge pre/post-test alongside a comparable non-IF control condition is justified once the perceived-quality requirements are met and a stimulus set implementing the design changes above is available.

## 5 CONCLUSIONS

Across a small STEM sample ($N$ = 22), scenarios from one pipeline configuration (open-weight LLM, single frozen prompt, one STEM sub-domain) were comprehensible and appropriately paced, engagement was marginal, and story–content coherence was the clear bottleneck — localized by the qualitative feedback in the narrative integration of the quiz, most prominently the artificial in-fiction motivation for quiz prompts and abrupt setting shifts. The study contributes a cautious first assessment of perceived quality for this pipeline configuration and a compact set of design implications for larger follow-up studies, including later work on learning effectiveness.

## ACKNOWLEDGEMENTS

The authors thank all study participants for their time and open feedback. This work was produced in the context of the project Life Labs, funded by Stiftung Innovation in der Hochschullehre within the funding programme Lehrarchitektur (grant number 1001-3214). *Author contributions:* F.R. led the work, including study design, implementation of the stimulus-generation and evaluation tooling, data collection and analysis, and original draft preparation. A.S. contributed to conceptualization, provided supervision, and performed substantial review and editing of the manuscript. Both authors have read and agreed to the submitted version of the manuscript.